# Extreme statistics, Gaussian statistics, and superdiffusion in global magnitude fluctuations in turbulence.

R. Labbé and G. Bustamante.

*Laboratorio de turbulencia, Departamento de Física, Facultad de Ciencia, Universidad de Santiago de Chile, USACH. Casilla 307, Correo 2, Santiago, Chile*

Extreme value statistics, or extreme statistics for short, refers to the statistics that characterizes rare events of either unusually high or low intensity: climate disasters like floods following extremely intense rains are among the principal examples. Extreme statistics is also found in fluctuations of global magnitudes in systems in thermal equilibrium, as well as in systems far from equilibrium. A remarkable example in this last class is fluctuations of injected power in confined turbulence. Here we report results in a confined von Kármán swirling flow, produced by two counter-rotating stirrers, in which quantities derived from the same global magnitude —the rotation rate of the stirrers— can display both, extreme and Gaussian statistics. On the one hand, we find that underlying the extreme statistics displayed by the global shear of the flow, there is a nearly Gaussian process resembling a white noise, corresponding to the action of the normal stresses exerted by the turbulent flow, integrated on the flow-driving surfaces of the stirrers. On the other hand, the magnitude displaying Gaussian statistics is the global rotation rate of the fluid, which happens to be a realization of a 1D diffusion where the variance of the angular increments $\theta(t+\Delta t)-\theta(t)$ scales as $\Delta t^\nu$, while the power spectral density of the angular speed follows a $1/f^\alpha$ scaling law. These scaling exponents are found to be $\alpha \approx 0.37$ and $\nu \approx 1.36$, which implies that this process can be described as a 1D superdiffusion.

# I. INTRODUCTION

Turbulence is a complex phenomenon: it is characterized by the interaction of structures in the flow velocity field distributed in ranges of space and time scales spanning several decades[1]. Thus, treating it analytically or by direct numerical computing remains a major challenge and performing experiments is, and possibly will last, a valuable research approach[2]. Since some 20 years to now, a number of experiments in confined turbulent flows have been performed in the so-called von Kármán swirling flow setup, in which a cylindrical vessel encloses a fluid stirred by two rotating disks with vanes, each facing the other and located near the ends of the container. One of the goals of these experiments is to increase the knowledge on energy transfer dynamics through turbulent flows, from the global scales to the smallest ones, where the energy is finally dissipated as heat. It was found that at the level of the energy injection scale, fluctuations of injected power are characterized by an extreme statistics: the probability density function (PDF) displays a non-Gaussian shape, strongly asymmetric with a stretched tail towards the low dissipation end[3,4]. The existence of similar PDFs in a variety of different systems raised considerable interest in fluctuations of global magnitudes with PDFs having this particular shape[5-7], linking together systems in thermal equilibrium with systems far from equilibrium, that is, those needing a permanent energy flow through them to stay in a statistically steady state. More recent works targeted the changes that under different conditions can undergo the statistics of global magnitudes in these systems[8-11]. Here we describe the results of a carefully controlled experiment in a confined von Kármán swirling flow, and show that extreme statistics, as well as Gaussian statistics, can be either displayed or hidden in global magnitudes derived from the same dynamical variables. Moreover, underlying the magnitude displaying extreme statistics we found that torque fluctuations have in fact a nearly Gaussian statistics, with small autocorrelation time and flat spectrum, suggesting that the torque resulting from the turbulent flow stresses acting on the surface of the stirring devices resembles a Gaussian white noise. In addition, we found that for the global magnitude having Gaussian statistics there is an underlying dynamics that can be described as a 1D superdiffusion process.

The remaining parts of this article are organized as follows: In Section II we define the experimental setup along with the magnitudes of interest, namely the angular speeds

$\Omega_1(t)$, $\Omega_2(t)$ of the stirrers and the torques $\tau_1(t)$, $\tau_2(t)$ acting on them. Then we give a brief deduction of the equations that govern the stirrers' motion by linearizing the dynamical equation for a rigid, symmetric rotor. This results in a system of two coupled Langevin equations, where the forcing terms are sums of torques applied by the electric motors, along with fluctuating torques arising from the hydrodynamic drag produced by the turbulent flow on the stirrers. We test this model by measuring the fluctuations of stirrers' angular speeds at two mean rotation rates: 16 rps (revolutions per second) and 32 rps. It happens that the onset of the fall-off of the stirrers' frequency response to turbulent fluctuations increases by the expected amount when the rotation rate doubles and the change in the kinematic viscosity due to the increase in the fluid temperature is taken into account, thus following the model prediction. Also, as expected, the stirrers behave like first order low-pass filters due to their inertia. This can be seen in the spectra of the angular speed fluctuations, which display a flat region in the low frequency range, and a $f^{-2}$ fall-off scaling past the first cutoff frequency, or a 6 dB/octave roll-off, as it is known in the electronics engineering literature.

In Section III the statistics of the sum of the stirrers' angular speeds is studied. Remembering that the stirrers rotate in opposite directions, the sum of these quantities corresponds to the global shear of the flow or, in other words, is proportional to the total power injected to the flow. Here we find that fluctuations of global power are consistent with asymmetric PDFs found in previous works, which is also the case for the fluctuations of the angular speed of each stirrer taken separately. By means of deconvolution of the angular speed signal, we retrieve the signal corresponding to the total torque acting on each stirrer. This procedure compensates only for the effects of stirrers inertia, and gives a flat spectrum for torque fluctuations that extends up to a second cutoff frequency. In fact, beyond that frequency the spectrum still displays a fall-off that scales as $f^{-2}$, again a 6dB/octave roll-off. As there is no reason for this secondary roll-off from the viewpoint of the stirrers dynamics, we ascribe this effect to the averaging of normal stresses on the flow-driving surfaces, assuming that the flow structures contributing to the total torque in the higher range of frequencies belong to a range of spatial scales smaller than the characteristic size of the vanes. Following this idea, we simply compensate the spectrum using a zero-phase-shift filter, which can be seen in the time domain as a non-causal filter, given that it use both, past and future events to deliver its output signal. In practice we implemented this filter in the frequency domain. As a result we obtain torque signal spectra that are flat in a four decade frequency band, with small autocorrelation times, and much more symmetric PDFs, having a nearly Gaussian shape.

In Section IV we consider the dynamics of the global rotation of the flow. Global flow rotation is simply related to the difference between the angular speeds of the stirrers, given that they rotate in opposite directions. We find that this global rotation displays some interesting scaling in the form of power laws. This happens for the power spectral density (PSD) of the global rotation angular speed $\Omega_R = \Omega_1 - \Omega_2$, where we observe that the PSD scales as $1/f^\alpha$ in the low frequency band of the spectrum. Given the symmetry of our experimental configuration, we should expect a symmetric PDF for $\Omega_R$. Although this is in fact the case, in addition, $\Omega_R$ happens to be a Gaussian variable, thus defining global flow rotation as a Gaussian process. This enables us to investigate the statistics of the global rotation angle, $\theta(t)$, defined as the time integral of $\Omega_R$. We find that in this case the variance of the increments of $\theta(t)$, defined as $\sigma^2(\theta) = \langle [\theta(t+\Delta t) - \theta(t)]^2 \rangle$, scales as $\Delta t^\nu$. Thus, the dynamics of the global flow rotation resembles a sort of 1D Brownian motion, but one in which the scaling laws correspond to what is known as an "anomalous" diffusion, as we will see later.

In the last section we provide some concluding remarks, where we attempt to give a coherent view of our findings, and relate them with results obtained in previous works. Finally, three appendixes provide for some technical details about the experimental setup, data reduction and analysis, and a numerical result confirming the consistency between the values for the exponents of the scaling laws obtained from the experiment.

## II. DEFINITION, DYNAMICS AND MEASUREMENT OF GLOBAL MAGNITUDES

Here, our focus will be set on a counter-rotating von Kármán swirling flow driven at constant torque, where one of the fluctuating magnitudes is the injected power

$$p = \Omega_1 \tau_1 + \Omega_2 \tau_2. \tag{1}$$

In equation (1) $\Omega_1$, $\Omega_2$, are the angular speeds of the rotating stirrers that drive the flow, and $\tau_1$, $\tau_2$, are the torques provided by electric motors that keep them rotating. In (1) we assume that torques related to mechanical and electrical losses are already discounted. Our experimental setup can be seen in figure 1. The turbulent flow is produced inside a cylindrical container by two low inertia stirrers, each one driven by a low inertia servomotor. These stirrers are located near the ends of the cylinder, and rotate in opposite directions. Details of the experimental setup are given in Appendix A. We made measurements at mean rotation rates of 16 rps and 32 rps. As usual, we define the Reynolds number for each stirrer as $\mathrm{Re} = \Omega R^2/\nu$, where $R$ is the radius of the stirrers, the angular speed, $\Omega$, is measured in rad s$^{-1}$, and $\nu$ is the kinematic viscosity of air. Thus, at the previously mentioned rotating rates, and having corrected the kinematic viscosity for each equilibrium temperature, the Reynolds numbers are $\mathrm{Re} = 2.1 \times 10^5$ at 32.5 °C, and $\mathrm{Re} = 3.4 \times 10^5$ at 72 °C, respectively. In equation (1), all of the magnitudes could be functions of time, but the analysis becomes simpler when some of them are held constant. Here we have chosen to keep torques constant, the so-called $\Gamma$ mode by Titon and Cadot[12]. This choice allows for a cleaner experiment: the risk of introducing artifacts because of controllers not optimally tuned, always present in experiments carried at constant velocity, is simply inexistent. This also allows for the stirrers to freely respond to changes in the flow reaction torque, contrarily to the case where their rotation rate is held constant. Ideally, this last condition would correspond to the case of stirrers having infinite inertia: a limit that is practically impossible to achieve, due to the natural limitations of servo controllers. Instead, in the constant torque mode only good quality voltage-controlled current sources are needed, along with motors capable of delivering constant torque at constant current, no matter their rotation rate. Both of these requirements are met by our experimental setup when it operates under the conditions already specified.

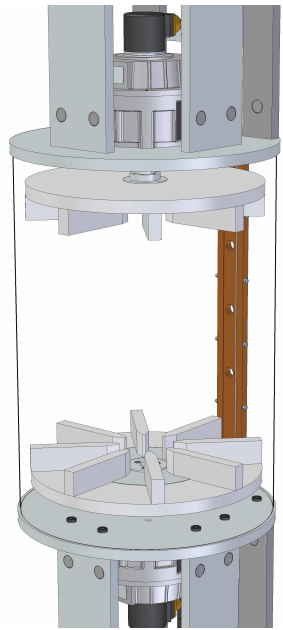

FIG. 1. Drawing of the experimental setup. Two low inertia stirrers are enclosed in a cylindrical vessel, each one driven by a low inertia servomotor (see text).

Now let us take a look to the equations of motion. Dimensional analysis states that the torque exerted by the turbulent flow on a rotating stirrer has the form $\tau_\mathrm{F} = -\eta_\mathrm{F} |\Omega| \Omega$, where $\eta_\mathrm{F} = C_\mathrm{D} \rho(P,T) R^5$. Here, $\rho$ is the density of the fluid, and is a function of the pressure $P$ and the absolute temperature $T$; $R$ is the radius of the stirrer, and $C_\mathrm{D}$ is a dimensionless coefficient that can be determined experimentally. Thus, keeping in mind that for each stirrer the sign of the angular velocity $\mathbf{\Omega}_i$ never reverses, and that we are writing these equations only for the magnitudes $\Omega_i$ of $\mathbf{\Omega}_i$, for $i = 1, 2$, we have

$$J \frac{\mathrm{d}\Omega_i}{\mathrm{d}t} + \gamma_\mathrm{M} \Omega_i + \eta_\mathrm{F} \Omega_i^2 = \tau_i^\mathrm{M} + \tilde{\tau}_i, \qquad (2)$$

where $J$ is the moment of inertia of each stirrer, including motor armature and rotating coupling components, $\gamma_M = 6.1 \times 10^{-4}$ Nms is the "viscous" coefficient for electrical losses in the motors, and is given by the manufacturer, $\tau_i^M$ is the constant torque delivered by the motor $i$, and $\tilde{\tau}_i$ is the (unknown) fluctuating torque exerted on the stirrer $i$ by the turbulent flow. As we are interested in the dynamics of fluctuations, we write the angular speed as a sum of a mean value, denoted by an over bar, and a fluctuating part, signaled by a tilde. Thus, assuming that the mean angular speed $\bar{\Omega}$ is the same for both stirrers, $\Omega_i = \bar{\Omega} + \tilde{\Omega}_i$. By inserting the previous expressions into equation (2), and noting that the mean flow torque on each stirrer, $\bar{\tau}_i^f = \gamma_M \bar{\Omega}_i + \eta_F \bar{\Omega}_i^2$, exactly compensates for the constant torque provided for its electric motor, after some little algebra and keeping only terms up to first order in fluctuating quantities we find that the stirrers' fluctuating motion is governed by the following pair of coupled Langevin equations:

$$J \frac{d\tilde{\Omega}_i}{dt} + \left(\gamma_M + 2\eta_F \bar{\Omega}\right) \tilde{\Omega}_i = \tilde{\tau}_i, \qquad i = 1, 2, \qquad (3)$$

As the mean value of the angular speed, $\bar{\Omega}$, was assumed identical for both stirrers, this implies a vanishing mean rotation rate for the global flow rotation. Thus, in (3) we neglected the weak torque exerted by the cylinder walls, although it is that torque what allows for the global rotation rate to have a definite mean value, which is nearly zero under the conditions of this experiment. We ascribe the coupling between the stirrers' fluctuating angular speeds $\tilde{\Omega}_1$ and $\tilde{\Omega}_2$ to the cross-correlation between the fluctuating torques $\tilde{\tau}_1$ and $\tilde{\tau}_2$. In the experiment, we obtained the quantities $\tilde{\Omega}_1$ and $\tilde{\Omega}_2$ from measurement of the angular speeds $\Omega_1$ and $\Omega_2$ by detrending the data using a quadratic polynomial. We did it that way because data trends related to slow drifts in the air temperature during the measurement runs are not always linear.

Equations (3) imply that there exists a cutoff frequency $\omega_c = 2\pi f_c$ for the fluctuations in the stirrers' angular speed,

$$f_c = \frac{\gamma_M + 2\eta_F \bar{\Omega}}{2\pi J}, \qquad (4)$$

which will be larger for faster mean angular speeds, and can be further increased if the total moment of inertia of the stirrer is reduced. In this experiment, we have $f_c^{(16)} \approx 0.525$ Hz and $f_c^{(32)} \approx 0.92$ Hz. Of course, we cannot reduce the inertia to zero, but using equations (3) the measured angular speeds can be deconvolved to obtain the torque exerted by the turbulent flow on the stirrers —a surrogate of a zero-inertia measurement— as we will see in the next section. Figure 2 shows the power spectral density (PSD) for angular speeds fluctuations at 16 rps and 32 rps. We note that the flat horizontal regions, along with the roll-offs displaying a $f^{-2}$ scaling past the first corner, are exactly the frequency responses associated to equations (3), suggesting that the spectrum of torque fluctuations remains flat beyond these first cutoff frequencies. This validates the idea of deconvolving the angular

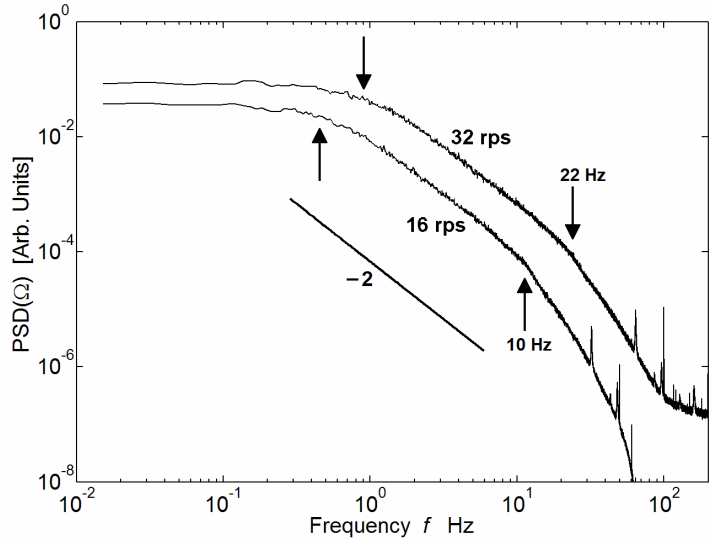

FIG. 2. Spectra of the angular speed $\Omega$ of the lower disk at two distinct angular speeds. The labels 16 rps and 32 rps indicate the mean rotation rate at which the corresponding spectrum was calculated. Arrows signal the corners where cutoff frequencies occur. Note the shift to the right of the cutoff frequencies when the rotation rate is increased from 16 rps to 32 rps. The bottom straight line has a slope $-2$ and represents an ideal spectrum that scales as $f^{-2}$, illustrating the effect of inertia on the angular speed fluctuations. The two rightmost arrows at roughly 10 Hz and 22 Hz signal cutoff corners not related to stirrers' inertia.

speed signal to obtain the signal corresponding to torque fluctuations.

The effect of the mean angular speed $\bar{\Omega}$ on the cutoff frequencies is evidenced by the shift to higher frequencies in the roll-off of the spectrum when the rotation rate is increased from 16 rps to 32 rps. The ratio between the spectra cutoff frequencies can be quantified by using equation (3), giving $f_c^{(32)}/f_c^{(16)} = 1.763$, whereas the ratio obtained from a fit of the frequency response of equations (3) to the measured spectra is $f_c^{(32)}/f_c^{(16)} = 1.752$. Note that the discrepancy is smaller than 0.7%, a remarkably good agreement suggesting that the approximation given by equations (3) adequately captures the global dynamics of the system.

In what follows, our task will be obtaining the fluctuating torques $\tilde{\tau}_i$ from fluctuations of the angular speeds, by deconvolving our experimental data for $\tilde{\Omega}_i$.

## III. GLOBAL SHEAR STATISTICS

If we take the signal corresponding to 32 rps and deconvolve it in the Fourier space, we arrive to the torque fluctuations, whose spectrum is labeled (D) in figure 3. In this process, by using appropriate filters we also removed all of the spurious peaks present in the primitive spectrum (see Appendix B). In spectrum (D) a second corner can be seen, which is also visible in the spectra labeled 16 rps and 32 rps in figure 2, at roughly 10 Hz and 22 Hz, respectively. These secondary cutoff frequencies are not related to inertia effects. They can be ascribed instead to the averaging of normal stresses exerted by the flow on the vertical surfaces of the vanes. We note that this cutoff is similar to that seen in a previous experiment in which the rotation speed was held constant[3], and that in both experiments they are located roughly at the same frequency. In that work inertia effects were strongly affected by the speed controller, and the observed cutoff —incorrectly attributed only to inertia— was indeed the combination of the second cutoff merged with an inertia cutoff shifted towards higher frequencies by the action of the servomechanism. This resulted in spectra displaying single corners, and single roll-offs regions for torque fluctu-ations with a $1/f^{-4}$ scaling beyond the cut-off frequency. In figure 3 this power law can be seen in the last region of the 32 rps spectrum, although in this case the pair of cutoff corners are well separated in frequency, as can also be seen in figure 2. This fact allows to clearly distinguish three almost linear regions in the spectra for angular speeds in figure 2, namely: the first, flat region, a second linear region scaling as $f^{-2}$, and the third region that scales as $f^{-4}$, as shown in the bottom spectrum of figure 3.

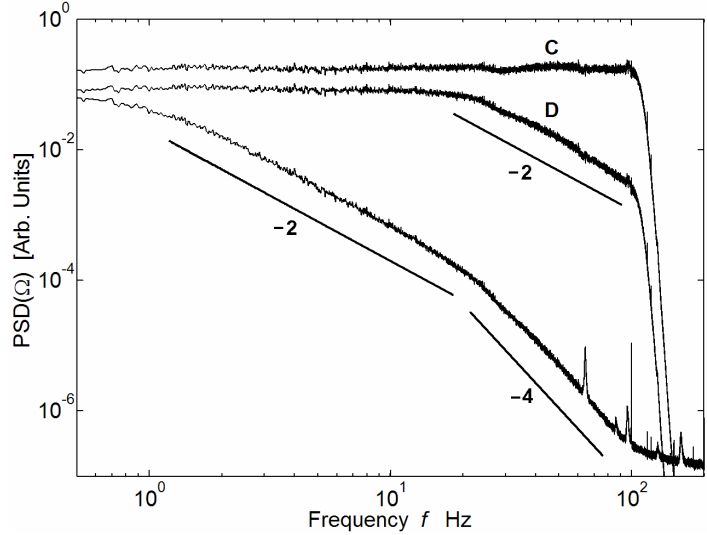

FIG. 3. The bottom curve is a detailed view of the roll-off regions of the spectrum labeled 32 rps in figure 2. The straight lines represent the (ideal) slopes in the corresponding regions of the experimental spectra. The spectrum labeled (D) results from deconvolution of the signal having the bottom spectrum. The flat spectrum labeled (C) and raised here by a factor of two, for clarity, results when the surface averaging is compensated (see text). Note that this spectrum spans four decades in frequency, given that it also contains the flat region of the spectrum labeled 32 rps in figure 2.

It is worth to stress here that the result obtained by deconvolution, which lead to the spectrum (D) in figure 3, is neither trivial nor tautological, because in the deconvolution process we used only information related to the dynamical equations that govern stirrers' motion. Thus, the spectrum labeled (D) corresponds indeed to fluctuations of torques which lead in turn to fluctuations of stirrers' angular speeds.

As we mentioned earlier, we can interpret the second cutoff frequency as the results of the surface averaging of structures in the turbulent velocity field with sizes smaller than the characteristic size of a vane. Many of these structures are advected by the mean flow and when they encounter the vanes, originate normal stresses that push the surface at more or less random times on different locations. Of course, smaller sizes and larger separation in these flow structures mean less correlated actions on the surface. The final effect of this process is a decreasing energy in the power spectrum of $\tilde{\Omega}$ with increasing frequency $f$, as higher frequencies are related to smaller and less correlated flow structures.

In principle, this averaging process could be deduced from the expression for the torque generated by the normal stresses acting on the vertical surface of the vanes. Assuming vanes of vanishing thickness, the starting equation should read

$$\boldsymbol{\tau}(t) = \int_S \mathbf{r} \times \boldsymbol{\sigma}(\mathbf{r},t) \cdot \hat{\mathbf{n}} \, \mathrm{d}S, \qquad (5)$$

were $\mathbf{r} = x_i \hat{\mathbf{e}}_i$ is a vector with its tail at some point on the rotation axis, and pointing to the surface element $\mathrm{d}S$, $\boldsymbol{\sigma}(\mathbf{r},t)$ is the stress tensor calculated in a reference frame attached to the stirrer, $\hat{\mathbf{n}}$ is the inward unit vector normal to the element $\mathrm{d}S$ on the surface of the vanes, $\hat{\mathbf{e}}_i$, $i = 1,2,3$, are the unit vectors along the coordinates axes, and $S$ is the union of all of the vertical surfaces of the vanes. The problem with equation (5) is that $\boldsymbol{\sigma}(\mathbf{r},t)$ must be calculated from the solution of the Navier-Stokes equations in a rotating frame that in addition undergoes angular acceleration. In Cartesian coordinates, assuming incompressible flow and appropriate initial and boundary conditions, the velocity field in this case must be solution of

$$\frac{\partial \mathbf{v}}{\partial t} + (\mathbf{v} \cdot \nabla)\mathbf{v} = -\frac{1}{\rho}\nabla p + \nu \nabla^2 \mathbf{v} - 2\boldsymbol{\Omega} \times \mathbf{v} - \boldsymbol{\Omega} \times (\boldsymbol{\Omega} \times \mathbf{r}) - \frac{\mathrm{d}\boldsymbol{\Omega}}{\mathrm{d}t} \times \mathbf{r}, \qquad (6)$$

where centrifugal, Coriolis and Euler forces are included. We think that this is a nightmare even for the best numerical solvers for turbulent flows, given that in our experiment the three last terms in equation (6) are certainly far from being negligible. Nevertheless, equation (6) is interesting because it implies that we should seek a self-consistent solution: this equation must be coupled to equations (3) to get $\boldsymbol{\Omega}$, but the driving terms $\tilde{\tau}_i$ need of the velocity field that solves (6) to evaluate (5), from which $\tilde{\tau}_i$, $i=1,2$ could be in turn calculated. For the moment, this seems to be a hopeless program, but equation (5) shows clearly the kind of surface integral involved in the production of the fluctuating torques $\tau_i$.

Leaving aside the previous considerations, we can try to compensate for the averaging effect using the following analogy: Inertia filtering is a sort of time average, one in which causality plays an important role. Only past events can enter in the average, and older events are forgotten with a weight factor that decreases with the elapsed time. This is related to the existence of a phase shift —which is a function of the frequency $f$— in the convolution of the time signal. We removed the inertia effect on the torque signal by using, in the frequency domain, the operator "inverse" to the differential operator acting on $\tilde{\Omega}_i$ in equations (3). Similarly, we can attempt to remove the effect of space averaging at the level of the time signal. The difference is that in this case there is no phase shift involved: the adding-up of signals is performed on the surface of the vanes. No causal relationship exists, at least in principle, between events occurring at different places on the surface of the vanes, although all of this is finally reflected in the time signal. Thus, we made a "deconvolution" process using a zero phase —non causal— time filter, designed to maintain a flat spectrum beyond the second cutoff corner because, from the flow dynamics viewpoint, as we previously signaled, rotation dynamics does not justify the second cutoff frequency displayed in figure 2. Unfortunately, in our analysis we cannot go beyond the frequency limit of 100 Hz that we have chosen, because the signal to noise ratio of our measurement at higher frequencies precludes it. In any case, as a result of the previously described procedure we obtained a signal with a flat spectrum that spans four decades in frequency, from about $10^{-2}$ Hz to $10^2$ Hz. In figure 3 this is displayed as the spectrum labeled (C), which we raised by a factor two to avoid the overlap with the other curves.

Now let us consider the PDFs of rotation rate fluctuations. In figure 4 the PDFs of fluctuations corresponding to the lower and upper stirrer are the dashed asymmetric curves. As we drive the stirrers at constant torque, from equation (1) we see that the instantaneous power injected into the flow is proportional to the sum $\Omega_1(t)+\Omega_2(t)$. The PDF of fluctuations of this magnitude is represented by the wider, continuous asymmetric curve, and given that the stirrers rotate in opposite directions, it corresponds to fluctuations in the global shear applied to the flow. The other global magnitude that we can derive from the angular speeds is the global rotation rate, given by the difference $\Omega_1(t) - \Omega_2(t) := \Omega_R(t)$. The PDF of this magnitude is displayed by the narrow symmetric curve, which happens to be *exactly* a Gaussian. To illustrate this, the inset is a representation using a logarithmic vertical scale, as usual, and the map $x \mapsto x|x|$ on the horizontal scale.

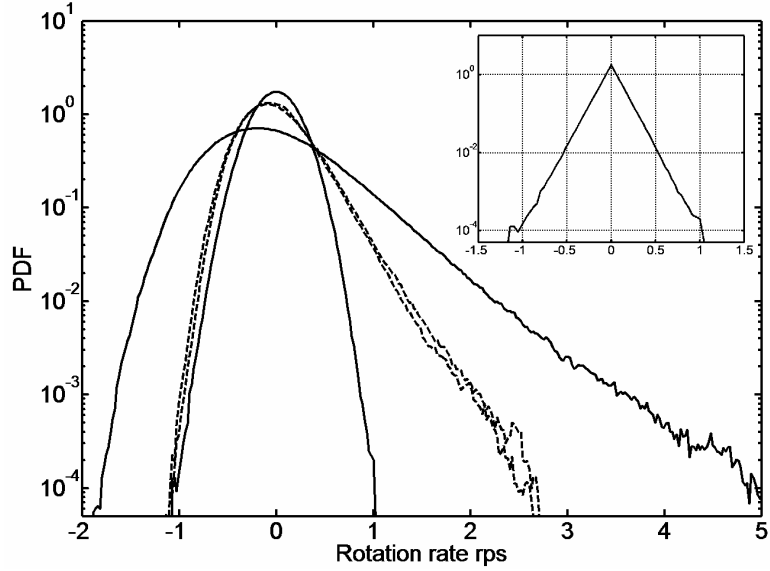

FIG. 4. Probability density functions. The asymmetric PDFs (dashed) correspond to fluctuations of the rotation rate in the lower and upper stirrers. The widest PDF corresponds to the sum $\tilde{\Omega}_1(t) + \tilde{\Omega}_2(t)$, which is proportional to fluctuations of the injected power. The narrow and symmetric PDF corresponds to the difference $\Omega_1(t) - \Omega_2(t)$, which is related to the global rotation performed by the flow. The inset represents this last PDF, where the map $x \mapsto x|x|$ was applied to the horizontal scale. This gives it an isosceles triangle-like shape, making evident its Gaussian character (see text).

Using these transformations a Gaussian becomes the sides of an isosceles triangle, which is just what we see in the inset. Thus, we have here a somehow surprising finding: On the one hand, small departures from the state of zero mean rotation, related to the difference $\Omega_1(t) - \Omega_2(t)$, display a Gaussian statistics. On the other hand, the injected power related to the sum $\Omega_1(t) + \Omega_2(t)$, displays an extreme statistics, with a PDF stretched towards high angular speeds. We note here that this last behavior is consistent with the result reported in the work mentioned previously[3]: at constant speed, the injected power displays intense events towards the low power end, corresponding to low torque events signaling a weakening of the drag exerted by the flow on the stirrers. In the present case the external torque applied to the stirrers is constant; hence a sudden weakening in the drag torque implies a sudden increase in the angular speed, which is reflected here in a PDF of $\tilde{\Omega}$ with a stretched right side.

Now we consider the effects on the PDFs of a) deconvolution; and b) compensation of surface averaging. The first procedure gives the PDF of torque fluctuations. The second one gives an estimate of torque fluctuations without surface averaging, in a range of scales corresponding to the frequency band from $10^{-2}$ Hz to $10^2$ Hz. In figure 5, the curve labeled (D) is the PDF of the torque. Note that the asymmetry is greatly reduced as compared to the PDF of the angular speed, but a stretching towards the right side still remains. As this is the result of compensating a sort of time averaging, we indeed expect an increase in the width of the PDF. In fact, this PDF is roughly five times wider than that for $\tilde{\Omega}$, as can be seen from the horizontal scale widening. It is worth remarking here that this is the primitive PDF from which an even more asymmetric PDF is obtained by averaging the underlying data. Nevertheless, this does not contradict the Central Limit Theorem, because the conditions of that theorem are no met by these global magnitudes. Indeed, time filtering due to the inertia of the stirrer increases the correlation time of the angular speed fluctuations.

Now, if we compensate for surface averaging, we get not only an even wider PDF but, in addition, a reversed left-right asymmetry, as displayed by the curve labeled (C) in figure 5. This gives to this PDF a shape reminiscent of that observed in the flow driven at constant angular speed in reference [3], but much closer to a Gaussian. Remembering that power fluctuations can be ascribed to fluctuations of global torque when angular speed is held constant, it appears that stirrers with less ability to respond to flow changes make normal stress fluctuations not only larger—as should be certainly expected—, but also make their statistics less Gaussian and even more asymmetric, a result which could be explained by the loss of local feedback from the flow into itself through the stirrer surface, because the response of the stirrer to flow stresses is highly reduced by the servo controller operating in the constant speed mode.

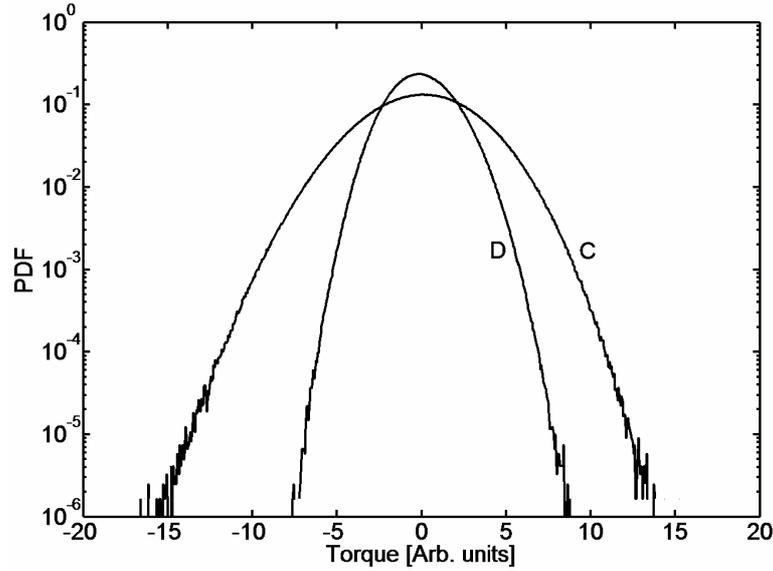

FIG. 5. Torque fluctuations PDFs, obtained from angular speed signals. The PDF labeled (D) is obtained after removing the effect of stirrer's inertia by deconvolution. The PDF labeled (C) corresponds to torque fluctuations after removing surface averaging from the time signal using a non-causal filter (see text). Note that the left-right asymmetry of curve (D) is reversed in (C).

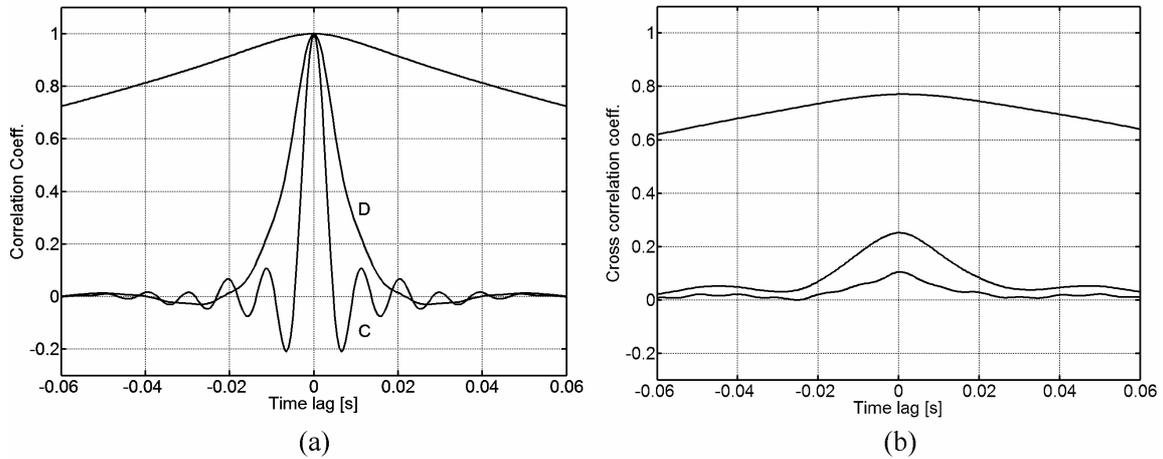

(a)    (b)

FIG. 6. (a) Autocorrelation functions of signals corresponding to spectra in figure 2. The wider curve corresponds to the spectrum labeled 32 rps in figure 2. The other two correspond to deconvolved (D) and compensated (C) signals. Autocorrelations fall to 0.2 in $2.43 \times 10^{-1}$ s (not shown), $1.12 \times 10^{-2}$ s, and $3.96 \times 10^{-3}$ s, respectively. (b) Cross correlation coefficients between upper and lower stirrers. The cross-correlation coefficient is strongly reduced, first by deconvolution (D) and then by compensation (C) of the signal (see text).

It is of interest to look at the time autocorrelation functions of fluctuations, which are displayed in figure 6 (a). The widest curve corresponds to the spectrum labeled 32 rps in figure 2, and the labels here indicate the correspondence with the deconvolved (D) and compensated (C) signals. It is clear that most of the correlation time for fluctuations in the angular speed comes from the stirrer inertia. When this effect is removed by means of deconvolution, the autocorrelation time decreases by a factor 22, as shown by curve (D), and by removing the surface averaging it is reduced further by a factor 2.8 (curve C). Thus, torque fluctuations with inertia and averaging effects removed display a flat spectrum spanning four decades, and an autocorrelation time on the order of four milliseconds. It is important to stress here that this time is only an upper bound to the autocorrelation time of torque fluctuations. In fact, the shape similar to a sinc function in curve (C) of figure 6 (a), suggest that this autocorrelation time is related to the frequency limit imposed by the low-pass filter that we used to remove the noise above 100 Hz. To test this idea we reduced by a factor two the bandwidth of this filter, which in fact approximately doubled the autocorrelation time. Thus, the autocorrelation time of four milliseconds is indeed related to the cutoff frequency of the filter used to remove the high frequency noise. If there was a possibility to increase the bandwidth of the measurements by improving the signal to noise ratio, this autocorrelation time would be even smaller than that displayed in curve (C) in figure 6 (a).

Finally, when the cross correlation between upper and lower stirrers is calculated, we see that inertia filtering has a major role in increasing the cross correlation coefficient. This is displayed in figure 5 (b) by the wider curve, whose maximum is $\chi_c \approx 0.75$. Removing inertia gives the curve labeled (D), with a maximum value of only $\chi_c = 0.25$. By removing the surface averaging effect we obtain the curve labeled (C), showing a cross correlation coefficient of only $\chi_c = 0.1$ at the maximum. This can be understood by considering that inertia filtering and surface averaging have the effect of partially removing uncorrelated "noise" coming from the main flow turbulence, which has the effect of decreasing the cross-correlation coefficient. Of course, a nonvanishing cross correlation between stirrers is to be expected, due to torque transmission from one stirrer to the other through the fluid motion. But curves (D) and (C) show that turbulent flow fluctuations are strong enough to mask up to about 90% of the cross-correlation between the stirrers' angular speed fluctuations.

## IV. GLOBAL ROTATION STATISTICS

When we consider the sum of equations (2), what we obtain is a new Langevin equation for fluctuations of a magnitude related to the global shear: $S(t) \sim \Omega_1(t) + \Omega_2(t)$, which displays a strongly asymmetric PDF, as we have seen. If we take instead the difference between equations (2), we get again a Langevin equation, but this time governing the global rotation of the flow. The PSD of $\Omega_R(t) = \Omega_1(t) - \Omega_2(t)$ is displayed in figure 7, for stirrers rotating at 16 rps and 32 rps. We see that below roughly 1.5 Hz and 2.6 Hz, respectively, the spectra labeled GR16 and GR32 are essentially straight lines, and follow a scaling law of the form $1/f^\alpha$ in that region. Thus, below those frequencies this flow performs a global rotation that fluctuates in a self-similar manner around the mean global angular speed with a power-law scaling. Above 1.5 Hz and 2.6 Hz, respectively, both spectra fall with increasingly steeper negative slopes, which indicates that global rotation is essentially a low frequency phenomenon, belonging to the global scale of the flow.

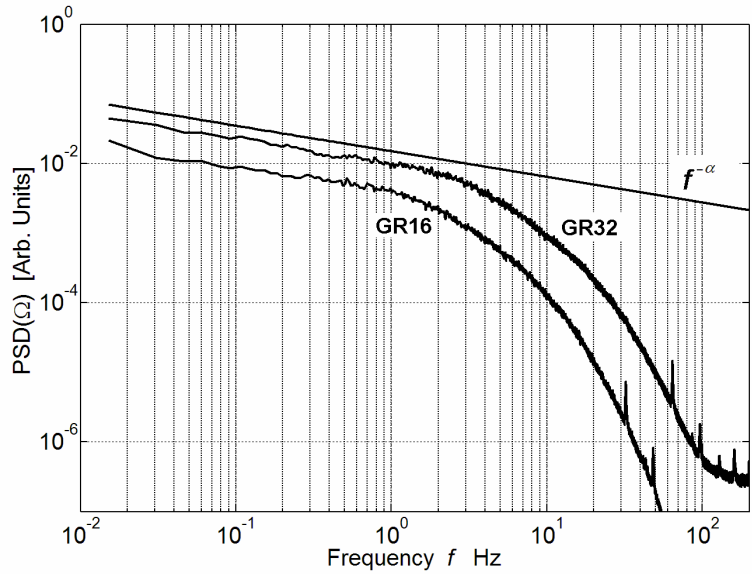

FIG. 7. Spectra of the signal $\Omega_R(t) = \Omega_1(t) - \Omega_2(t)$, corresponding to the global rotation of the flow. The straight line with scaling $1/f^\alpha$ looks parallel to the straight, low frequency bands of the spectra GR16 and GR32 (see text).

It turns out that the value of the exponent is approximately $\alpha = -0.37$, which is the slope of the straight line in figure 7 (see Appendix C). Note that the spectrum for stirrers rotating at 16 rps (the bottom curve in figure 7) shares the same features, except for the shift of its cutoff corner to a lower frequency, of about 1.5 Hz. Remembering from figure 4 that the PDF of $\Omega_R(t)$ is Gaussian, we conclude here that global rotation is a Gaussian process characterized by a Gaussian-distributed function $\Omega_R(t)$ which, in the relevant frequency interval, has a spectrum that scales closely as $f^{-0.37}$. Given that we are dealing with an angular speed, it is natural to look at the flow global angle, which we define as

$$\theta(t) = \int_0^t \Omega_R(\zeta)d\zeta \approx \theta_k = \sum_{k=0}^n \Omega_R(t_i)\Delta t, \qquad (7)$$

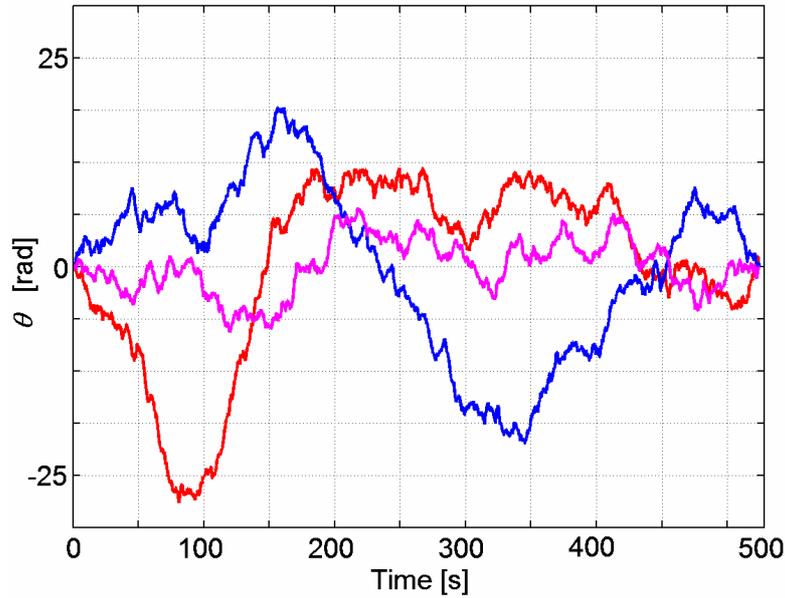

FIG. 8. (Color online). Examples of sample-paths, corresponding to the integral of three samples of the global rotation angular speed. As they start and finish at the same value (zero in this case), these paths are known as bridges.

were $t_k = k\Delta t$ is the discretized time, and $\Delta t$ is the time interval between sample points, for $0 \leq t \leq T$ and $0 \leq n \leq N$. In our experiment, we took samples consisting of $N = 2^{20}$ points in a time $T \approx 524$ s. We used an oversampling factor $\beta \approx 4$, which makes the approximation made in (7) good enough[13]. For each one of these samples, equation (7) defines a *sample-path*, in the language of Brownian motion theory. Given that the mean value of $\Omega_R(t)$ is zero, these paths are in fact *bridges*[14, 15]: they start and finish at the same value; in this case, at $\theta = 0$. Figure 8 displays three examples of these sample-paths, or bridges.

The next question that we can ask is: what is the scaling of the variance of the increments of the global rotation? The response that we found from our data, in an interval of time increments compatible with the spectrum GR32 in figure 8, is

$$\left\langle \left[ \theta(t+\Delta t) - \theta(t) \right]^2 \right\rangle = C \Delta t^{\nu}, \tag{8}$$

where $C$ is a constant and $\nu \approx 1.36$ (see Appendix C). Being this a Gaussian process, it follows from the scaling (8) that the structure functions introduced by Kolmogorov in 1941[16,17], applied to time increments, must have the following scaling:

$$\left\langle \left| \theta(t+\Delta t) - \theta(t) \right|^n \right\rangle = C_n \Delta t^{n\nu/2}, . \tag{9}$$

The dotted curves in figure 9 display the left hand side of equation (9), obtained from the experimental data. In the same figure the straight lines correspond to the r.h.s. of equation (6), for $n = 2,\ldots,7$ and $\nu \approx 1.36$, corresponding to the exponent found for the variance. For clarity, the straight lines in this figure were raised by a little amount above the dots, and the vertical separation of structure functions was increased by factors of ten. In reality, all of these curves have intersections in points in a neighborhood of $(2.95, 0.2)$.

It can be seen that the slopes of the straight lines appear to match properly the slopes of the straight portion of the experimental points, in an interval spanning approximately one decade in the time increments: between $0.25$ s and $3.3$ s, which is compatible with the frequency band where the spectrum GR32 in figure 8 follows a power law. Thus, the global rotation exhibits a behavior compatible with an anomalous diffusion, specifically a superdiffusion[18] following scaling laws where the exponent of the variance of the increments is $\nu = 1.358 \pm 0.002$, accordingly with the linear and logarithmic fits given in Table I of Appendix C for $n = 2$.

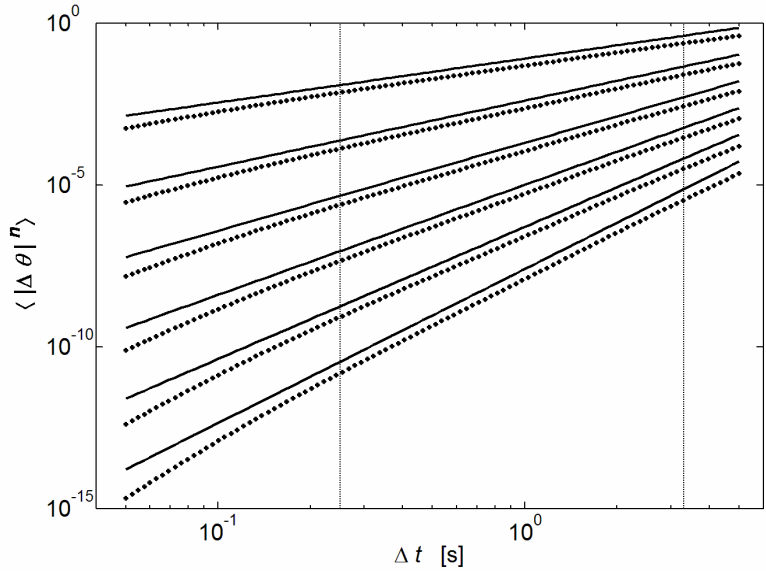

FIG. 9. Dots represent Kolmogorov time structure functions calculated from the experimental data. The straight lines have slopes corresponding to the exponent of time increments in equation (9). Fits of exponents were performed in the region between the vertical lines. From top to bottom, $n = 2, \ldots, 7$. (see text).

It is worth to stress here that the scaling laws we have found for the global rotation phenomenon are restricted to low frequencies, and that in the case of the spectrum GR32 these frequencies are below $2.6$ Hz. Although the self-similar region of this spectrum spans only two decades, it could be extended to even lower frequencies by performing measurements for periods longer than $T \approx 524$ s, which is the duration of each one of the sample records in the experiments reported here. In this way, the corresponding structure functions should have a power law region for time increments longer than the upper limit $\Delta t = 3.3$ s displayed in figure 9. With respect to the lower limit $\Delta t = 0.25$ s, the spectra in figure 7 show that it depends on the rotation rate of the stirrers. In principle, by increasing this rate it would be possible to extend the power law region of the structure functions towards shorter time increments. Nevertheless, there will be always a limit imposed by the loss of correlation between the stirrers beyond the frequencies where the global rotation spectrum is falling.

**V. CONCLUDING REMARKS**

Most of the observed extreme statistics in fluctuations of global magnitudes in von Kármán swirling flows that we report here appears to be related to the dynamics of the stirring devices, or to the averaging of normal stresses on the vertical surface of vanes. Interestingly, the extreme statistics observed at constant rotation speed[3] seems to be linked to a rather extreme driving mode: the one corresponding to stirrers having nearly infinite moment of inertia. In addition to the increase in the asymmetry of the PDF, the constant speed mode should increase the rms amplitude of torque fluctuations. This last conclusion is also given in a work in which this system was modeled by means of a stochastic Langevin equation[19]. Another interesting observation is that, once inertia and surface averaging are removed, it is found that torque fluctuations seem to behave like a white noise, having an almost Gaussian statistics[20], a small autocorrelation time, and a flat spectrum. Given that in the experiments by Titon and Cadot[12] water was used as working fluid, our conjecture to explain the difference between their almost symmetric rotation rate PDFs, as compared with those obtained in air (when the flow is driven with stirrers rotating at constant torque), is that such difference might be related to the huge ratio (~$10^3$) between the densities of water and air. In air, even using our light stirrers, inertia has a central role in the rotation dynamics. On the contrary, normal stresses exerted by water at similar Reynolds numbers will be roughly a hundred times larger. Thus, a stirrer driven at constant torque will be far more easily forced to follow the fluctuations in time of the space-averaged rotation of water near its surface. Additionally, this means that a servo motor will have a hard time attempting to keep a constant rotation rate in water if its power is similar to that used in experiments performed in air. An experiment in which geometrical and dynamical similitude is maintained in setups working in water and air is underway, to test whether angular speed PDFs are indeed different if water is used instead air.

Perhaps our most remarkable observation is the contrast between the statistics of global rotation versus global power injection (or global shear): when the driving device is allowed to respond to forces arising from the flow motion, the injected power fluctuations display a highly asymmetric PDF, having a right side with a stretched exponential shape, whereas global rotation fluctuations, derived from the same magnitudes, namely the stirrers' angular speed fluctuations, have a perfectly Gaussian PDF. This last process is found to be a realization of a superdiffusion[18] by the motion of the turbulent flow at the global scale level. In Appendix C we give a numerical result that reasonably supports that if the spectrum exponent is $\alpha \approx 0.37$, then the variance scales with an exponent $\nu \approx 1.36$.

We find indeed remarkable that the dynamics of global rotation in this turbulent flow, a system that we can properly classify as being *complex*, could be described by a far more simple 1D superdiffusion: an anomalous version of a 1D Brownian motion, with spectra and structure functions following power laws. This last behavior is always of interest in a number of fields in science, and finding a turbulent fluid motion process that can be described by power laws in such a simple manner is, in our view, indeed remarkable. From a practical point of view, we believe that these results might be of interest to validate turbulence models used in numerical simulation of confined turbulent flows.

## ACKNOWLEDGMENTS

We gratefully acknowledge S. Fauve, L. Vergara and I. Ispolatov for critical reading of the manuscript and suggestions. R.L. is indebted to G. Bobadilla, S. Navarro and G. Palma, whose support made possible this work as part of a sabbatical year project, and to P. Umbanhowar for highlighting, a long time ago, the advantages of pancake servomotors. Financial support for this work was provided by FONDECYT under grants #1090686 and #1040291.

## APPENDIX A: MATERIALS AND METHODS

The experimental setup used in this work is similar to that used in previous works[3,4], except by the size and the main components of the stirrers, which here were made of expanded polystyrene to minimize their moment of inertia. Located near the ends of a cylindrical container of 59.5 cm height and 50.5 cm diameter, each one is driven by a low inertia servomotor. The cylindrical wall of the container, shown transparent in figure 1, is made of a thin aluminium sheet with a thickness of 0.5 mm, painted with flat-black paint on the external face to facilitate radiant heat transfer to the ambient at high temperatures. This allows a temperature reduction of about 10% at temperatures reaching 100 ºC. External air cooling provided by a fan allows some additional 15% reduction of heating. The torque applied to each servomotor is controlled by a voltage controlled current source, and the instantaneous rotation rates of the stirrers are measured using two high resolution optical encoders. The diameter of the stirrers is $D = 36$ cm and their separation is $H \approx 47$ cm. The disks and their 8 vanes were cut from an expanded polystyrene plate of thickness $T = 1.5$ cm and density $\rho = 0.02$ gcm$^3$. They were assembled and covered with a thin layer of polyvinyl acetate glue, to increase their stiffness. Reinforcements made of thin aluminium disks of thickness $t = 0.05$ cm and diameter $d = 15$ cm were added in the central zone. These assemblies were directly coupled to Kollmorguen Servodisk® model JR16M4CH, low inertia, pancake type servomotors, having a coefficient of viscous losses $\gamma_M = 6.1 \times 10^{-4}$ Nms. The resulting moment of inertia of each stirrer, comprising polystyrene disk and vanes; motor armature; shaft; iron and aluminium coupling and reinforcing parts; and bolts is $J \approx 4.7 \times 10^{-3} \,(\pm 10\%)$ kgm$^2$. In comparison, a stirrer having the same size but made of three-layer plywood of 6 mm thickness and driven by a 1 kW DC motor having an iron-core armature, would have a moment of inertia $J \approx 5 \times 10^{-2}$ kgm$^2$, which is about 10 times larger. When these stirrers rotate in opposite directions at approximately the same rotation rate, we found that the non dimensional coefficient in the expression $\eta_F = C_D \rho(P,T) R^5$ is $C_D = 0.204$. This is obtained by a least squares fit to $\tau_F = -\eta_F \Omega^2$, taking in consideration the variation of the air properties with the temperature. The electric motors were driven by two Kollmorguen pulse width modulated (PWM) servo amplifiers type KXA-175 used in torque mode, meaning that in this case they operate merely as voltage-controlled current sources. The stirrers' angular speeds $\Omega_i$ were measured by means of frequency to voltage converters, using the pulses delivered by two Kollmorgen BA25I optical encoders attached to the shafts of the electric motors. These encoders give 1000 pulses per turn, and two of the output channels, A and B, are in phase quadrature. By using custom circuitry to detect transitions instead levels, a resolution of

4000 pulses per revolution was obtained. All physical quantities were linearly mapped to voltages in the 0 to 10 V interval. Specifically, the absolute stirrer rotation rate can vary between 0 and 60 rps. This range was mapped to the interval $[0, 10]$ V. The absolute values of torque delivered by the DC motors vary from 0 to 3.72 Nm, which was also mapped to the $[0, 10]$ V interval. The experiment control was performed by a personal computer running programs under the National Instruments LabView® platform. Electric signals were low-pass filtered using a IOtech Filter488/8 eight channel, $8^{th}$ order elliptic low-pass filter, programmed through a National Instruments AT-GPIB/TNT board, and digitized by a National Instruments AT-MIO-16X multifunction board. Data were acquired at 2 kS/s sampling rate using a cuttoff frequency of 250 Hz. This oversampling level allows an excellent representation of signals up to the highest frequency passed by the filter. Whenever possible, appropriate offsets and gains were applied to the signals to take full advantage of the 16 bit digitizer card resolution.

**APPENDIX B: DATA REDUCTION AND ANALYSIS**

Appropriate level of shielding was used for signal cabling, but some hum from the power grid at 50 Hz and multiples is always present. This can be seen in some of the spectra in figure 1 (e.g. the spectrum labeled 32 rps). Also, some peaks at several other frequencies, related to vibrations at the disks rotation frequencies and multiples, which can be ascribed to stirrers unbalancing, and angular vibrations related to electrical asymmetries in the servomotors are present. All in all, our data allows for a maximum spectral frequency of 1 kHz, but no relevant signal components exist beyond 100 Hz. Thus, to clean up signals we implemented a number of filters in the Fourier domain. All the spectral components beyond 120 Hz were removed by using a low-pass, zero-phase filter having a reversed sigmoid shape, with 0 dB gain in the $[0, 100]$ Hz band and an attenuation higher than 350 dB for $f > 120$ Hz. The frequency response of this filter is given by

$$H_N(f) = \frac{1}{2}\left\{1 - \tanh\left[s\left(|f| - f_c\right)\right]\right\}, \quad \text{(B1)}$$

where $s$ is a steepness factor. In our case, $s = 0.1$. Grid hum components were removed with zero-phase, Lorentzian-like band-stop filters tuned at $50 \text{ Hz}$ and $100 \text{ Hz}$. These filters have the following frequency response:

$$H_{\text{hum}}(f) = 1 - \frac{g_n \zeta}{\left(|f| - f_0\right)^\nu + \zeta}, \tag{B2}$$

where the parameters $g_n$, $\zeta$ and $\nu$ were adjusted to match the shape of the hum peaks. Peaks related to vibrations and electrical asymmetries have their physical origin in the system dynamics. Causal filters are appropriate in this case, and we found that *LRC*-like notch filters work fine to remove these peaks. Their transfer function (in the Laplace domain) is

$$G(s) = \frac{s^2 + (k/\tau_r)s + \omega_0^2}{s^2 + (1/\tau_r)s + \omega_0^2}. \tag{B3}$$

Here, $\omega_0$ and $\tau_r$ are the center frequency and relaxation time, respectively, whereas $0 < k < 1$ determines the attenuation.

Filtering is also appropriate to deconvolve inertia filtering in torque fluctuations. Given that the stirrers are governed by a first order ordinary differential equation, the dynamical effect on the torque signal is simply that of a first order low-pass filter, having a transfer function

$$G_\Omega(s) = \frac{1}{\tau_\Omega s + 1}, \tag{B4}$$

where $\tau_\Omega$ is the relaxation time. Its value is calculated from equation (3), which gives

$$\tau_J = \frac{J}{\gamma_M + 2\eta_F \bar{\bar{\Omega}}}. \tag{B5}$$

Deconvolution in Fourier domain is achieved simply by dividing the angular speed signal by $G_J(i\omega)$ in the Fourier domain.

At this point, a remark is in order: At constant torque drive mode, the drag torque signal obtained by deconvolution is not necessarily equivalent to the torque signal obtained when measurements are performed with stirrers rotating at constant angular speed. As we pointed out in the main text, this case ideally corresponds to stirrers having infinite inertia. Thus, there is little angular acceleration of the stirrer in response to normal stresses on the vertical surfaces of the vanes, which implies that there is almost no local normal stress feedback from one place to another in the neighborhood of the stirrer surface. As a consequence, we should expect some differences in the flow dynamics. This should explain why by deconvolving the signal we do not obtain PDFs with the shape of those obtained at constant angular speed.

**APPENDIX C: ESTIMATES OF THE SCALING EXPONENTS**

It is known that determining exponents of power law distributions by standard methods is likely to produce biased estimates[21-23]. This is especially true when absolute values of exponents in power laws are larger than one. In our case, we have a spectrum with a power law scaling $1/f^\alpha$ spanning two decades with $\alpha \approx 0.37$, and a power law scaling for the variance of the increments with an exponent $\nu \approx 1.36$, which appears to be valid in a little more than one decade of time increments. To support our statements about the values of these exponents, we estimated independently their values using both, logarithmic and linear fits for $\alpha$, and for each one of the structure functions. Given that $\alpha < 0.5$, the PSD values span something less than one decade within the frequency range where the power law is valid. Then, using least squares fits in both, logarithmic and linear scales makes sense. We perform these fits in the frequency interval from $7.6 \times 10^{-3}$ Hz to $2$ Hz, which contains $262$ spectrum points. From the logarithmic fit we obtain

$$\alpha = 0.367, \qquad \text{(C1)}$$

with $R^2=0.931$ and 95% confidence bounds $(0.354, 0.379)$.

Performing a fit using

$$S = Af^{-\alpha}, \tag{C2}$$

where $S$ is the PSD, we obtain

$$\alpha = 0.370, \tag{C3}$$

with $R^2=0.97$ and 95% confidence bounds $(0.364, 0.376)$.

Now we consider the structure functions exponents. The scaling exponents are given by

$$\zeta_n = \frac{n\nu}{2}, \quad n = 2,3,\ldots, \tag{C4}$$

and we can proceed by fitting straight lines to the data for each $n$, and from each exponent we can find a value for $\nu$, using both linear and logarithmic scale fits. We summarize our findings in Table I.

TABLE I. Values of $\nu$ extracted from the structure functions exponents.

| | LINEAR SCALE FIT | | LOG SCALE FIT | |
|---|---|---|---|---|
| $n$ | $\nu$ | 95% confidence bounds | $\nu$ | 95% confidence bounds |
| 2 | 1.3590 | 1.3580, 1.3600 | 1.3570 | 1.3550, 1.3600 |
| 3 | 1.3506 | 1.3494, 1.3520 | 1.3586 | 1.3566, 1.3614 |
| 4 | 1.3525 | 1.3510, 1.3535 | 1.3605 | 1.3580, 1.3630 |
| 5 | 1.3524 | 1.3512, 1.3536 | 1.3620 | 1.3600, 1.3644 |
| 6 | 1.3520 | 1.3510, 1.3526 | 1.3637 | 1.3614, 1.3660 |
| 7 | 1.3692 | 1.3683, 1.3697 | 1.3652 | 1.3632, 1.3675 |

As we can see, the values of $\nu$ in Table I do not differ too much. At the very least, we have two coincident digits in all of the cases, and coincidences of three digits can be seen in somewhat more than 50% of them. We see a slight increasing trend in the logarithmic scale fits which does not seem to appear in the linear scale fits.

To conclude this analysis, we give here the result of a numerical calculation using a synthetic signal $s_j$, $j = 1, 2, \ldots, N$, consisting of 40 samples, each one having $N = 2^{22}$ points, giving a total of $M \approx 1.7 \times 10^8$ sample points, and having a spectrum that scales as $1/f^\alpha$, with $\alpha = 0.37$. The spectrum is plotted in figure 10 (a). The 2$^{nd}$ order structure function (or variance of time increments) of the discrete time integral of $s$ is displayed in figure 10 (b). It scales as $\Delta t^\nu$, with $\nu = 1.36$. The agreement is very good, showing that the values of the exponents found in the experiment are consistent with the result of the numerical calculation. In other words, a superdiffusion where the spectrum of the velocity increments scales as $1/f^\alpha$, with $\alpha \approx 0.37$, is characterized by a variance of time increments following a power law scaling with exponent $\nu \approx 1.36$.

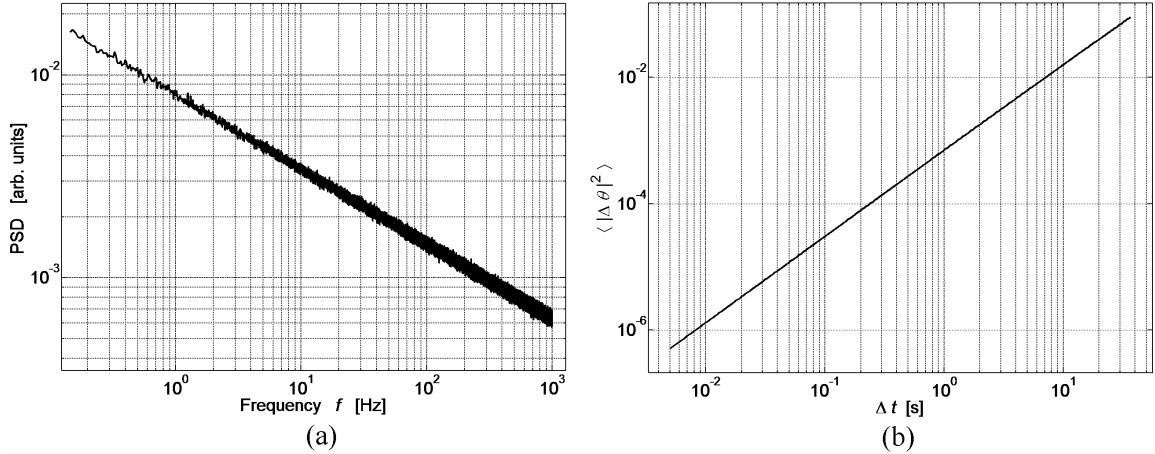

FIG. 10. (a) Spectrum of a synthetic signal scaling as $1/f^\alpha$ with $\alpha = 0.37$. (b) The corresponding 2$^{nd}$ order structure function, with scaling exponent $\nu = 1.36$.